\documentclass{article}
\usepackage{spconf,amsmath,graphicx}
\usepackage{blindtext}
\usepackage{enumitem}
\usepackage{amssymb}
\usepackage[]{algorithm2e}
\usepackage{amssymb}
\usepackage{caption}
\usepackage{graphicx}
\usepackage{subcaption}
\usepackage{xcolor}
\usepackage{placeins}

\newcommand{\xb}{{\textbf{x}}}

\newcommand{\ab}{{\textbf{a}}}

\title{Distributed interference cancellation in multi-agent scenarios}
\name{Mahdi Shamsi, Alireza Moslemi Haghighi, Farokh Marvasti}
\address{Multimedia and Signal processing Lab (MSL),
	Advanced Communications Research Institute (ACRI),\\
	EE Department of Sharif University of Technology, Tehran, I.R.Iran}
\begin{document}
%

\maketitle
\begin{abstract}
This paper considers the problem of detecting impaired and noisy nodes over network.  
In a distributed algorithm, lots of processing units are incorporating and communicating with each other to reach a global goal.
Due to each one's state in the shared environment, they can help the other nodes or mislead them (due to noise or a deliberate attempt).
Previous works mainly focused on proper locating agents and weight assignment based on initial environment state to minimize malfunctioning of noisy nodes.
We propose an algorithm to be able to adapt sharing weights according to behavior of the agents. Applying the introduced algorithm to a multi-agent RL scenario and the well-known diffusion LMS demonstrates its capability and generality.
\end{abstract}
\section{Introduction}
\label{sec:format}
In this correspondence, the problem of error propagation in a distributed algorithm is investigated.
There is a variety of examples in which distributed algorithms are used such as distributed estimation\cite{cattivelli2009diffusion,zayyani2016distributed}, sensor networks\cite{liu2016sensor}, distributed control systems\cite{ge2017distributed,zhang2017distributed}, swarm intelligence algorithms\cite{hassanien2018swarm,blum2015swarm}, multi-agent reinforcement learning scenarios\cite{foerster2018counterfactual,lowe2017multi} and signal processing on graph\cite{chen2015discrete,ortega2018graph,perraudin2017stationary} just to name but a few.
Considering the general viewpoint of graph over networks; each algorithm in its  core consists of nodes(agents) as the computing units and links as the communication path.
It should be noted that distributed algorithms can provide us with robustness, reliability and faster convergence in exchange for a larger number of low-cost processing units.
A simplifying assumption, for developing and analyzing distributed algorithms, is considering the shared environment to be akin to all of the graph nodes\cite{mohammadi2019square}.
Assuming a stationary behavior of an environment, several algorithms have been proposed for optimizing, locating and placing a node in the initial state. There are also certain ways of analyzing the environment for assigning weights to each node, Since there could be shortcomings due to relying on each node's observed environment.
However, in a non-stationary scenario, the environmental behavior, and hence performance of the nodes, can be changed. This could be under the influence of different factors, like noise increasing 
or an intervening procedure such as attacking and hacking.

This article is organized as follows.
In Section \ref{sec:Nprop}, we study on noise propagation in three typical scenarios over a graph. We begin with a discrete probabilistic non-linear state model and then simplify it to a linear sharing framework. 
 Then we propose an algorithm for malfunctioning detection in a distributed scenario. 
 Section \ref{sec:DistEst} is dedicated to distributed estimation; the Adapt Then Combine (ATC) scenario is studied as a prominent approach for diffusion Least Mean Square (LMS)\cite{chen2015diffusion,abdolee2016diffusion}. Capability of the proposed method is illustrated by applying to the ATC with a high noisy node in different situations [of dense and sparse connection graph].
In Section \ref{sec:MA-RL}, a manipulated RL environment is introduced to be as a multi-agent approach for studying an intervened scenario.
At last, Section \ref{sec:conclusion} concludes the paper.
\section{Noise Propagation}
\label{sec:Nprop}
In this section, we introduce error propagation as a general issue in distributed algorithms.
For the purpose of generality, some simplifications are applied to this procedure.
From the control systems modeling perspective, all the distributed contents and functionalities are stated by overall system's state equations.
Every node can be assumed an agent or an estimator. Hence the connecting links are interpreted as corresponding connections and sharing procedures.
In this point of view,
we are not only interested in system's steady state. The transient behavior of nodes are also a major concern.
In this system, it is assumed that there are N nodes which are in corporation with each other to reach a global goal.
In a general case, all of A, B, C and D are nonlinear operators.
A general state equations are shown as
\begin{align*}
\vec{x}(n)=A(\vec{x}(n))+B(\vec{u}(n));\,
\vec{y}(n)=C(\vec{x}(n))+D(\vec{u}(n)).
\end{align*}
By assuming a graph as a distributed platform, functionality of each can be analyzed in terms of signal flow graph paradigm.
This assumption leads us to interpret each operator more clearly.
Hence, with minor alterations, the operator A becomes the well-known adjacency matrix of the main graph.
For the purpose of simplicity, these operators are assumed to be linear. Furthermore, it is sufficient to study information sharing procedures, which are commonly linear like in recent distributed estimation methods\cite{cattivelli2009diffusion}.
\begin{align*}
\vec{x}(n)=\tilde{A}.\vec{x}(n)+B.\vec{u}(n);\;\vec{\tilde{\textbf{a}}}_i=\frac{\vec{\textbf{a}}_i}{mean(\vec{\textbf{a}}_i)}
\end{align*} 
where $\vec{\tilde{\textbf{a}}}$ and $\vec{\textbf{a}}_i$ are the $i^{th}$rows of $\tilde{A}$ and $A$ (Adjacency matrix).
Another simplification that we should mention is that we are using a symmetric directional graph. It means, each pair of connected nodes can communicate with each other. However, the effect of the shared value of each node in the destination node is determined by the number of incoming links to the latter one.
Generally, in combination step of sharing procedures of distributed algorithms, all incoming estimated values are averaged. Thus, sum of elements in each row of A should be normalized to 1.

As a desirable feature of distributed structure, at least in many cases, isomorphism of the main graph is considered in the following way; each two nodes can be connected independent of the others and all nodes have the same level of importance.
There are trends focusing on optimizing significance of nodes based on their processing power or environmental turbulence\cite{mohammadi2019square,zabini2016random}. These approaches mainly concentrate on proper locating of the agents\cite{udwadia1994methodology}; in other words, optimizing the placement procedure.

Hence, by assuming an arbitrary node labeled as $i$, all other nodes have the same expected data flow from $i$.
Due to the isomorphism of the graph, with malfunctioning of an arbitrary node (which could be due to the high level of noise or hacking attacks), all the other nodes are supposed to receive the same level of wrong (misleading) data on average.
It is clear that, these methods consider a stationary condition. Hence, after proceeding through certain number of steps of the main function of the system, it seems reasonable to think of the present environmental state.

For a better comprehension, and without loss of generality, we only consider the case of a single impaired agent. It can be easily extended to the larger number of damaged nodes because of the simplifying linearity assumption.
It means we can study two arbitrary nodes in the graph and analyze the procedure of exchanging data. In order to predict the incoming data from the node $i$ to the arbitrary destination node $j$, we should know expected number of paths between them.
It can be done through three different cases.
By defining $p \triangleq\frac{M-1}{N-2}$ and $s\triangleq \frac{M}{N-1}$ where N and M are respectively number of all nodes and average number of neighbors of an arbitrary node, we have 3 cases as shown in the following table.
\begin{table}[h]
	\begin{tabular}{|c|c|c|}
		\hline
		$i$ & case          & $\eta_{_{K> 1}}^{^{<i>}}$ \scriptsize{= expected \#paths of length} $K$                                     \\ \hline
		$1$ & no-loops      & $P(N-2,k-1)\times s\times p^{k-1}$                              \\ \hline
		$2$ & no self-loops & $(N-2)^{K-1}\times s^{K-1}\times p$                              \\ \hline
		$3$ & general       & $\sum_{\ell=0}^{K-1}(N-2)^{K-\ell-1}\times s^{K-\ell-1}\times p$ \\ \hline
	\end{tabular}
	\label{tabel:paths}
	\caption{Expected \#paths of length $K$ between two arbitrary nodes [= $\eta_{_{K}}^{^{<i>}}$], in a graph with $N$ nodes and, on average, $M$ neighbors for each one; for all $3$ cases: $\eta_{_{1}}^{^{<i>}}=s$.}
\end{table}
In the $2^{nd}$ case, the paths are supposed to be determined at the destination node.


It is claimed that, the third case can be interpreted as the impulse response of the overall system between two arbitrary nodes of the pre-described graph. Hence, one node can introduce a pandemic interference over the graph.
It must be mentioned that, even in simple distributed methods, each node is continuously compensating the incoming distortion based on its predefined functionality. However we are insisting on their limited capability of compensation.
In order to prevent noise propagation in a graph, we introduce a simple method of weighting and successor decision making on node elimination.
In each step, each node receive some data from its neighbors (including itself). Using a simple procedure of weighting it will then use all of the incoming estimations thoroughly; in which, a node attempts to reach a predefined goal via optimizing a cost function (the goal can also be non-stationary).
Although different ordering of these two steps results in various algorithms, it would not harm the main body of our proposed algorithm.
Additionally, we can assume a malfunctioning node in some general graph. In the combination step, each node can determine the distance between its own estimation and its neighbors' shared estimations. Based on these distances, it can assign a proper weight to each one.
Hence it can eliminate detected malfunctioning nodes if it seems necessary.
It can also put some sharing restrictions on the detected impaired nodes to isolate them from following the righteous agents. It can start dissembling and after being trusted, do its job (just as an example of smart simple  hacking).
The proposed algorithm is shown as in Alg.\ref{alg:detectionAlg}.
\begin{algorithm}

		\For{$t=1,\dots,T$}
		{\For{$i \in \{1,\dots,N\}$}{Adaptation: $\tilde{\xb}_i(t) = g(\xb_i(t-1),\theta_i(t))$\\
		Combination: $\xb_i(t)=h(\tilde{\xb}_j(t)\vert j\in N_i)$}}
	\caption{The Algorithm of Malfunctioning Detection: $g(.)$ and $h(.)$ are respectively adaption and combination operators; $T$ and $N$ are the number of iterations and the total number of nodes; $x_i(t)$, $\tilde{x_i}(t)$, $\theta_i(t)$ and $N_i$ are the estimate of the $i^{th}$ node, its temporary version, its insight from the sharing environment at time $t$ and the set of its Neighbors.}
	\label{alg:detectionAlg}
\end{algorithm}
Based on the main target algorithm, the distance calculating function, the weighting method, and the eliminating strategy can be properly defined.
\section{Distributed Estimation}
\label{sec:DistEst}
In this section, we illustrate the capability of our proposed algorithm for detecting noisy estimator in a distributed estimation method. We assume well-known ATC strategy\cite{tu2012diffusion}. The main algorithm, with the respect to Alg.\ref{alg:detectionAlg}, can be summarized as follows\cite{tu2012diffusion}
\begin{equation}
g(\xb_i(t-1), \theta_i(t))= \Bigg\{\begin{array}{cc}
d_i(t)\leftarrow&\ab_{i,t}^T\xb_{opt}+\nu_i(t)\\
e_i(t)\leftarrow&d_i(t)-\ab_{i,t}^T\xb_i(t-1)\\
\tilde{\xb}_i(t)=&\xb_i(t-1)-\mu_{i,t}e_i(t)\ab_{i,t}\notag
\end{array}
\end{equation}
\begin{equation}
h(\tilde{\xb}_j(t)\vert j\in N_i)=\Bigg\{\begin{array}{cc}
\alpha_{i,j}(t)\leftarrow
(\frac{\|\tilde{\xb}_i(t)-\tilde{\xb}_j(t)\|_2}{\zeta})^e\\
\forall j\in N_i: c_{i,j}(t)\leftarrow
 \frac{exp(\alpha_{i,j}(t))}{\sum_{k\in N_i}c_{l,i}exp(\alpha_{i,k}(t))}\\
\xb_{i}(t)=\sum_{l\in N_i}c_{i,l}\tilde{\xb}_l(t)\notag
\end{array}
\end{equation}
where for the $i^{th}$ node at time instance $t$.
\begin{table}[h]
	\begin{tabular}{|c|c|c|}
		\hline
		$\ab_{i,t}$               & measurement vector                        & description      \\ \hline
		$\nu_i(t)$                & measurement noise                         & variance: $\sigma_i^2$                                                                                          \\ \hline
		$\mu_{i,t}$               & adaptation rate                           & step-size                                                                                                       \\ \hline
		\multicolumn{1}{|l|}{$\zeta$} & \multicolumn{1}{l|}{normalization coefficient} & related to $\sigma_i$                                                                                           \\ \hline
		\multicolumn{1}{|l|}{$e$} & \multicolumn{1}{l|}{determining decision type}       & \begin{tabular}[c]{@{}l@{}}uniform: $e = 0$\\ hard: $e\rightarrow\infty$\\ soft: $e\in (0 ,\infty)$\end{tabular} \\ \hline
	\end{tabular}
\label{tabel:param}
\end{table}
It is an implementation of diffusion LMS algorithm which has given an outstanding performance\cite{tu2012diffusion}.
It was mentioned that in each iteration, we can have an adaptation step and a combination step. However, we should be cautious, since with low value of adaptation rate, possibly all of the estimators in each neighborhood have near estimations.
Hence, simply feeding the distance calculator with their values can mislead the total procedure.

\begin{figure}[h]
	\centering
	\begin{subfigure}[b]{0.23\textwidth}
		\centering
		\includegraphics[width=\textwidth]{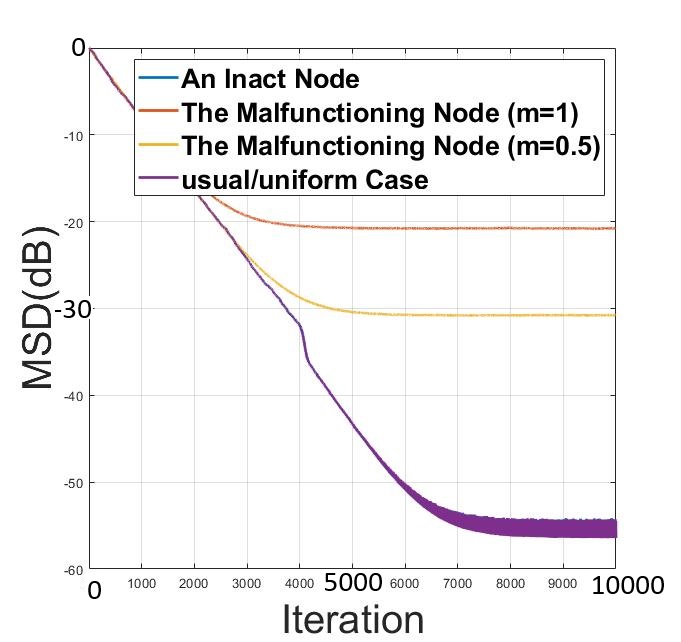}
		\caption{\scriptsize{$\rho=0.1, \sigma_{noise}=0.04, N=50, \zeta=0.015, e=2$.}}
		\label{fig:fig1}
	\end{subfigure}
	\hfill
	\begin{subfigure}[b]{0.23\textwidth}
		\centering
		\includegraphics[width=\textwidth]{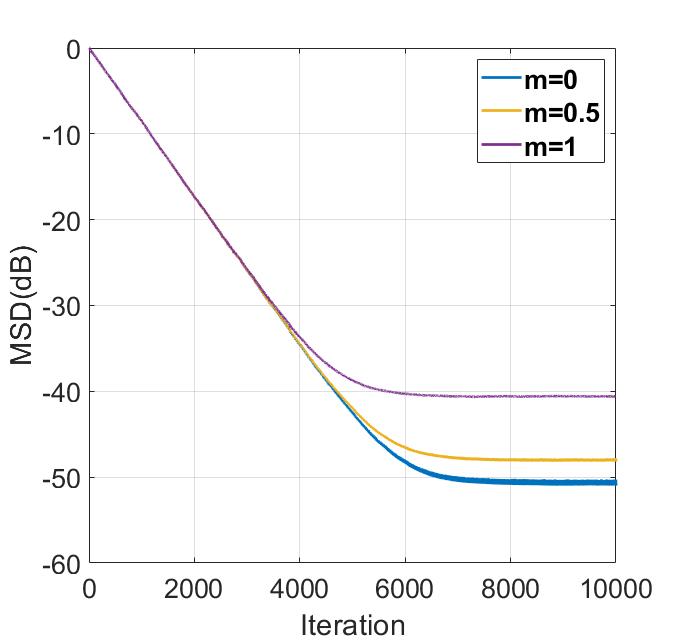}
		\caption{\scriptsize{Inact nodes connected to an impaired node with uniform weights, $\rho=1, \sigma_{noise}=0.04, N=10, e=0$.}}
		\label{fig:fig2}
	\end{subfigure}
	\hfill
	\begin{subfigure}[b]{0.23\textwidth}
		\centering
		\includegraphics[width=\textwidth]{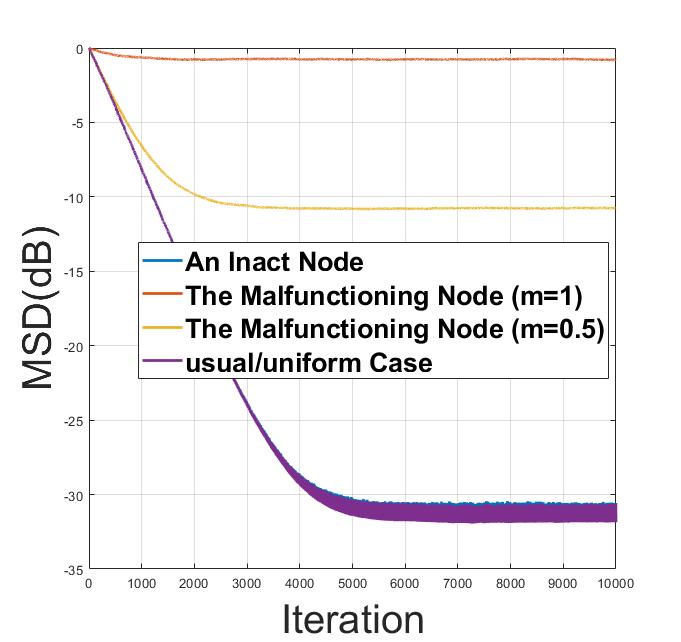}
		\caption{\scriptsize{$\rho=0.2, \sigma_{noise}=0.4, N=100, \zeta=0.05, e=8$.}}
		\label{fig:fig3}
	\end{subfigure}
\hfill
\begin{subfigure}[b]{0.23\textwidth}
\centering
\includegraphics[width=\textwidth]{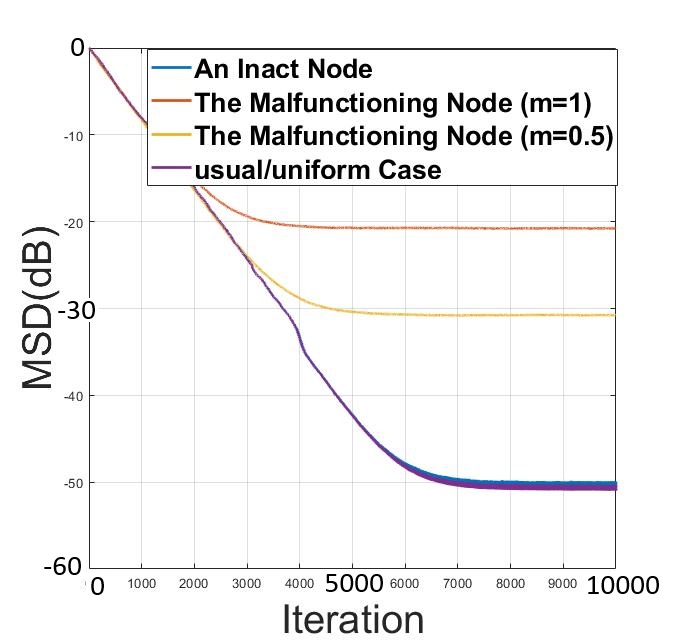}
\caption{\scriptsize{$\rho=1, \sigma_{noise}=0.04, N=10, \zeta=0.015, e=8$.}}
\label{fig:fig4}
\end{subfigure}
	\caption{Distributed Estimation Performance (ATC scenario), MSD vs \#iteration; $L=100, s=0.5, \mu=0.001, T_a=10$, $\sigma_{impaired}=10^m\times\sigma_{noise}, \#simulation=1000$.}
	\label{fig:DE}
\end{figure}

In order to simulate the proposed algorithm, we assume a graph having N nodes with s as an existence probability of a link between each pair of nodes ($\rho$). The target signal has a length of L and is generated with normal distribution (with  given sparsity of nonzero components: $s$).
In fact, we just extend the adaptation time ($T_a$).
In order to resolve this issue, each node starts its own simple adaptation step multiple times. After an agreed adaptation time slot, sharing method is run.
As shown in Fig.\ref{fig:DE}, the heavily-noisy estimator is dislodged from the combination step. In the both cases of dense and sparse graph, improvement is clear.
Since the noisy one still has access to the intact estimations, it can follow the estimation's trend, as the fluctuations can be seen in the Mean Square Deviation (MSD) curves. By tuning the weighting parameters ($\zeta$ and $m$), it can be prohibited. 
\section{Multi-Agent RL}
\label{sec:MA-RL}
In reinforcement learning scenario we will rely on some other mathematical foundations to be helped with, which in the following paragraphs it will be explored in more details. In this case, exploring the search space while exploiting on the best strategy which has been found so far, will give the so-called agent the ability to adapt to its environment \cite{sutton2018reinforcement}. 
We will consider the classic RL framework in which an agent is interacting with a finite state Markovian environment on a discrete time scale \cite{bellemare2017distributional}. Our final goal will be maximizing the expected of the utility function $Q$ over all possible policies, considering different $R(s,a)$ rewards for each pair of state-actions \cite{still2012information}. Bellman's equation describes the following in terms of both, the expected reward and expected outcome of the random transition\cite{bellemare2017distributional}. The notation of $\pi$ and $\gamma$ allude respectively to the certain policy and the discount factor\cite{sutton2018reinforcement}.
%

Based on the optimality equation(\ref{eq:optEq}), several algorithms have been proposed\cite{bellemare2017distributional}, which in this setting we are going to focus on one of the well-known methods, $Q$-Learning \cite{sutton2018reinforcement,bellemare2017distributional},
as shown by \ref{eq:Qlearn}
{\small{
	\begin{align}
&Q^*={\mathop{\mathbb{E}}}[R(s,a)]+
\gamma{\mathop{\mathbb{E}}}^{P}[\max_{\{a'\in A\}}Q^*(s',a')]
\label{eq:optEq}\\
&Q^{new}(s_t,a_t)\leftarrow(1-\alpha)Q(s_t,a_t)\dots\notag\\
&+\alpha(R(s_t,a_t)
+\gamma\max_{\{a'\in A\}}Q(S_{t+1}',a'))
\label{eq:Qlearn}
\end{align}
}}

In the case of finite state-action pairs we use the tool of Q-table for our convenience. Q-table is a simple representation of the Q-values in each pair of state-action. Since, in the beginning of the process our initialized Q-table acts randomly, our total regret \cite{osband2017deep} diverges linearly from its optimal state. Thus, to overcome this issue, another modification is to use an $\epsilon$-decay exploration rate \cite{osband2017deep}. In this case, our initial exploration rate will be reduced through the process of learning. 
\begin{align}
\scriptsize{\epsilon=\epsilon_{min}+(\epsilon_{max}-\epsilon_{min})\times \mathrm{e}^{-\#step\times DecayRate}\notag}
\end{align}
In a multi-agent setting, we rely on multiple agents for completing a given task. All of the above definitions could be extended for larger number of agents. There is just an extra point that here agents will interact with one another. Considering the case of fully-connected network, all agents will participate for filling the Q-table. Passing the Q-values, will give them an access to their local neighborhoods. As a result, we expect to converge faster than before. In a simple case, each agent would fill its own Q-table with averaging on all observations they have been given. Finally, after a large number of episodes we may expect to see that their Q-tables have converged to a single Q-table, which could be used for all of them afterwards.   

\subsection{Problem and Model definition:}
Given the uncertainties of the real-world problems we should always be prepared for some obstacles. For instance, in the case of multi-agent learning, one could mention the broken state of an agent as a critical situation. Since they are all transferring information and using those information for their actions, a major turbulence could be misleading. So, we can think of some adjustments applied to the prior methods. Here we are going to detect those broken agents and rooting them out from our decision-making, by a sort of swarm-intelligence algorithm. In this way of analysis, agents themselves will decide on that which agents are broken. Consider an arbitrary agent $i$ and its neighborhood as $N_i$, in the case of fully-connected network. Then for each $j\in N_i$, in addition to what their given observation is, we would also get their fakeness indices as well. Finally, agent $i$ will decide based on its given information that which agent(s) is (are) malfunctioning. Here, we just consider the simple case of one broken agent. However, the results could be generalized to more complex situations. Furthermore, we would also assume that our algorithm will converge which, results in finding the correct target. 
\begin{equation}
	f_j\triangleq \text{Fakeness labels from nearby agents}\notag
\end{equation}
\begin{equation}
	f_i^0=\Bigg\{\begin{array}{cc}
			k=arg\max_{\{j\in N_i\}}(Q_j-Q_i)\;\; &;Q_j-Q_i>0\\
			None &;o.w\notag
			\end{array}
\end{equation}
Then we do the majority vote. So, first for each agent in the system s, we count its appearance as a broken (fake) agent. Finally, we decide whether our maximum appearance should be reported, based on comparing its votes against the total number of votes.
\begin{equation}
\scriptsize {\forall n\in S;\;n_i=1{_{(n=f_i^0)}}+\sum_{j\in N_i}1_{\{n=f_j\}}\notag}
\end{equation}
\begin{equation}
\scriptsize	{f_i=\Bigg\{\begin{array}{cc}
k=arg\max_{\{n\in S\}}(n_i)\;\; &;k_i>\frac{\sum_{j\in N_i}1}{2}\\
None &;o.w\notag
\end{array}}
\end{equation}
Again, due to the randomness being involved in the beginning of the learning, we do not adjust our weights right after a single step. Instead, we wait for $10$ steps as an example and then decide based on those $10$ steps that which agents were behaving badly. This would help us to avoid random decisions and to reduce some weights unexpectedly. In order to average our observations we would also normalize them by reducing the weight of the broken agent linearly by the factor of $0<\lambda<1$ It should be kept in mind that we consider the agent itself as one of its neighbors as well.
\begin{equation}
\scriptsize{	g({{\textbf{Q}}}_i^{m-1}, \theta_i(t))=\\
	\Bigg\{\begin{array}{cc}
	r\leftarrow R(s_t,a_t)+\gamma\max_{\{a'\in A\}}Q_i^{m-1}(s_{t+1},a')\\
	\tilde{{{\textbf{Q}}}}^m_i(s_t,a_t)=(1-\alpha)Q_i^{m-1}(s_{t},a_t)+\alpha r\notag
	\end{array}}
\end{equation}
\begin{equation}
	\scriptsize{h(\tilde{{{\textbf{Q}}}}^m_i(s_t,a_t)|j\in N_i)=\\
	\Bigg\{\begin{array}{ccc}
	&f_i\leftarrow[fakeness\, voting]\\
	
	&c_{i,f_i}(m)\leftarrow\frac{1-\lambda}{\sum_{j\in N_i}1}; \,\text{if:}\,f_i\in N_i\\
	
	&c_{i,j}(m)\leftarrow\frac{1+\frac{\lambda}{-1+\sum_{j\in N_i}1}}{\sum_{j\in N_i}1}; \,\forall f_i\in N_i\setminus f_i \\
	&{{\textbf{Q}}}_i^m(s_t,a_t)=\sum_{l\in N_i}c_{i,l}(m)\tilde{{{\textbf{Q}}}}^m_j(s_t,a_t)\notag
	\end{array}}
	\end{equation}
For comparing our results, we have used the OpenAI gym package in python $3.7.2$. The SnowFlake environment, with $8\times 8$ grid is our main basis. We have also manipulated the environment for putting different agents in different starting points. As it could be seen from the table below, our impaired diagnosis algorithm outperforms the classical approach in which we ignore the impaired agents. It should be noted that here we have the simple case of one broken agent, which is detected correctly. We also consider a task to be done whenever there is an agent which has made it to the target point.
\begin{table}[h]
	\centering
	\scriptsize{\begin{tabular}{|c|c|c|c|}
		\hline
		Parameter       		 & case 1	& case 2& case 3\\ \hline
		Number of episodes       & $1e6$	& $1e5$	& $1e6$ \\ \hline
		Maximum number of steps  & 1000		& 10000	& 1000  \\ \hline
		Learning Rate            & 0.8		& 0.8	& 0.7	\\ \hline
		Discount Factor          & 0.97     & 0.97  & 0.97  \\ \hline
		Minimum Exploration Rate & 0.001	& 0.001 & 0.001 \\ \hline
		Maximum Exploration Rate & 1     	& 1     & 1     \\ \hline
		Exploration Decay Rate   & 0.001 	& 0.001 & 0.001 \\ \hline
		Number of Agents         & 3     	& 3     & 3     \\ \hline
		$\lambda$                & 0.7   	& 0.9   & 0.7	\\ \hline
success rate (with detection)    & \%84  	&\%78   & \%71	\\ \hline
success rate (without detection) & \%53  	& \%59  & \%55	\\ \hline
	\end{tabular}}
\end{table}
\section{Conclusion}
\label{sec:conclusion}
By emphasizing on the role of non-stationarity in an interacting network, we introduced a novel approach for analyzing it. The main focus of our analysis was also on the problem of error diagnosis and canceling certain interferences caused by damaged agents in network. We also applied our proposed algorithm to two different frameworks with the same issue of an impaired node. In both cases of distributed estimation and multi-agent RL, by detecting and reducing the effect of the impaired node, we surpassed our prior performance.    
\vfill\pagebreak
\bibliographystyle{IEEEbib}
\bibliography{bibfile}

\end{document}